# Specific heat study of $Ga_{1-x}Mn_xAs$

Sh. U. Yuldashev [a], Kh. T. Igamberdiev, Sejoon Lee, Y. H. Kwon, and T. W. Kang
Quantum-Functional Semiconductor Research Center, Dongguk University, Seoul 100-715, Korea

Yongmin Kim and Hyunsik Im
Department of Semiconductor Science, Dongguk University, Seoul 100-715, Korea

A. G. Shashkov
Institute of Heat and Mass Transfer, 15P. Brouki, Minsk 220072, Belarus

Specific heat measurements were used to study the magnetic phase transition in $Ga_{1-x}Mn_xAs$. Two different types of $Ga_{1-x}Mn_xAs$ samples have been investigated. The sample with a Mn concentration of 1.6% shows insulating behavior, and the sample with a Mn concentration of 2.6% is metallic. The temperature dependence of the specific heat for both samples reveals a pronounced λ shaped peak near the Curie temperature, which indicates a second-order phase transition in these samples. The critical behavior of the specific heat for $Ga_{1-x}Mn_xAs$ samples is consistent with the mean-field behavior with Gaussian fluctuations of the magnetization in the close vicinity of $T_C$.

[a]E-mail: shavkat@dongguk.edu

The ferromagnetic $Ga_{1-x}Mn_xAs$ semiconductors have been studied intensely over the last decade and have become a model system for diluted magnetic semiconductors.[1-3] It is now widely accepted that the ferromagnetism in $Ga_{1-x}Mn_xAs$ arises from the hole-mediated interaction between the Mn local magnetic moments.[3] However, over the past years, an intense debate has sparked about the nature of the hole states in this material. Whether the holes are free and reside in a valence band[4] or they are localized in an impurity band?[5] The study of the critical behavior of ferromagnets near the Curie temperature is very helpful in understanding their magnetic and electronic properties. Establishing the universality class for the phase transition provides information on the range of the exchange interactions. For example, a long-range exchange interaction in the case of mean-field and a short-range value in the case of Heisenberg or Ising models.[6] Very recently, the Curie point singularity in the temperature derivative of resistivity in $Ga_{1-x}Mn_xAs$ with nominal Mn concentration ranging from 4.5 % to 12.5 % has been investigated.[7] Using the similarity between the critical behaviors of the *dρ/d* and the specific heat for metallic ferromagnets,[8] the critical exponent α of the specific heat has been estimated from the log( *dρ/dt)* vs log(t) plots. All data sets collapsed into the common temperature dependence for $T < T_C$, as well as another common dependence was observed for $T > T_C$. However, no clear power-law behavior in *dρ/dT* on the either side of the transition has been observed.

In this paper, we present the results on experimental study of the specific heat of $Ga_{1-x}Mn_xAs$ with a relatively low concentration of Mn near the Curie temperature. The temperature dependencies of resistivity, magnetization and specific heat have been conducted. To the best of our knowledge no specific heat measurement results for $Ga_{1-x}Mn_xAs$ have been reported so far. The $Ga_{1-x}Mn_xAs$ layers with a low Mn concentrations ($x < 0.03$) were grown on semi-insulating (001) GaAs substrates by using MBE. The epilayers with thickness about 1μm were grown at low temperature of 270 $^0$C with different temperatures of the Mn source. The Mn concentration in the layers was estimated from x-ray diffraction measurements and it was additionally confirmed by x-ray microanalysis. No post-growth thermal annealing was performed. The hole concentration measured at room temperature for the two typical samples with the concentration of Mn (1.6% and 2.6%) were of $2.7 \times 10^{19}$ cm$^{-3}$ and $4.5 \times 10^{19}$ cm$^{-3}$, respectively. We are aware that the concentration of free holes obtained from Hall measurements for $Ga_{1-x}Mn_xAs$ is not rigorously correct because of the presence of terms arising from the anomalous Hall effect (AHE). However, because of the relatively low Mn concentration in the samples studied, the AHE contribution is expected to be negligibly small at room temperature. The temperature dependence of the resistivity was conducted by using LakeShore system equipped with a closed cycle cooling cryostat. The temperature dependence of magnetization was measured by a superconducting quantum interference device (SQUID) magnetometer. The specific heat was measured by using a 3ω method described elsewhere.[9] The heating current frequency was of 1 MHz and the sweep rate was of 100 mK/min.

Figure 1 shows the temperature dependence of the magnetization for the two of $Ga_{1-x}Mn_xAs$ samples with the Mn concentration of 1.6% (sample A) and 2.6% (sample B). The measurements were conducted at the magnetic field of 10 Oe applied parallel to the sample plane after cooling of samples in a zero magnetic field. From magnetization curves the Curie temperatures of these samples have been determined and the respective $T_C$ are shown in Fig.1 by solid arrows. The Curie temperature for samples A and B is about 40 K and 52K, respectively. The temperature dependence of the resisitivity for these samples, measured at zero magnetic field, is shown in the inset of Fig.1. It is seen that the sample A demonstrates an insulating behavior, while the sample B shows a metallic behavior. Both samples exhibited a maximum (a rounded cusp) at the Curie temperature $T_C$, marked by solid arrows. It should be noted that the resistivity maximum coincides with the Curie temperature, determined from the magnetization curves, within the experimental errors. In our previous paper[10] the resistivity maximum at the Curie temperature of $Ga_{1-x}Mn_xAs$ was explained by the magneto-impurity model proposed by Nagaev.[11]

Figure 2 shows the temperature dependence of the specific heat $C_p$ for $Ga_{1-x}Mn_xAs$ samples (solid lines) and GaAs substrate (dash-dotted line). The specific heat curves of $Ga_{1-x}Mn_xAs$ samples show a pronounced λ shaped peak, which indicates an existence of a second-order phase transition in these samples. The specific peak maximum in $Ga_{1-x}Mn_xAs$ samples located near the Curie temperature and therefore, it is attributed to the ferromagnetic-paramagnetic phase transition. As is seen in Fig. 2, with increasing of the manganese concentration the specific heat peak increases in the amplitude and shifts to higher temperature. The specific heat peak maximum for the $Ga_{1-x}Mn_xAs$ samples A and B was observed at 39.95 K and 51.75 K, respectively.

Figures 3 and 4 show the temperature dependence of the magnetic specific heat $C_{mag}$ for the samples investigated, which was obtained by subtracting off a smooth background of the specific heat of GaAs substrate. The nonmagnetic contribution of $Ga_{1-x}Mn_xAs$ layers to the specific heat is supposed to be very close to the specific heat of GaAs because the Mn concentration in the samples investigated is relatively low. The critical behavior of the specific heat near the phase transition is described by $C_p = C^{\pm} t^{-\alpha}$, where $C^{\pm}$ are the critical amplitudes of the specific heat above (+) and below (-) $T_C$, $t = |T-T_C| / T_C$ is the reduced temperature, and α is the critical exponent of specific heat.[12] The insets of Figs. 3 and 4 show the plots of the magnetic specific heat versus the reduced temperature in a double logarithmic scale, where $T_C$ is the maximum of the specific heat peak. It is seen, that for $10^{-3} \leq t \leq 10^{-2}$ of the reduced temperature interval close to the $T_C$, the experimental data above and below $T_C$ have a similar slope. For sample A the slope is about 0.09, while for sample B the slope is about 0.5. The value of the critical exponent $\alpha = 0.09$ is close to the critical exponent $\alpha \approx 0.1$ of the three-dimensional (3D) Ising model. This is an unexpected result because the Ising critical behavior is valid for a short-range exchange interaction, while a long-range mean-field-like exchange interaction is expected. However, near the second–order ferromagnetic phase transition, not only the specific heat $C_p$ shows power-law dependence on the reduced temperature, but other parameters such as spontaneous magnetization, magnetic susceptibility reveal the similar behavior with critical exponents β and γ, respectively, and at $T_C$ $M(H) \propto H^{1/\delta}$.[12] Figure 5 shows the M(H) curve for sample A measured at the Curie temperature of 40K. The Arrot plot of $M^2$ vs H/M for this sample is shown in the inset (a) of Fig.5. It is seen that the isothermal curve is linear, which demonstrates the mean-field behavior with $\beta \approx 0.5$ and $\gamma \approx 1$. In order to determine δ, we plot M vs H in a double logarithmic scale shown in the inset (b) of Fig.5. The inverse slope of log M vs log H gives $\delta = 3.1$ which is close to the mean-field value of 3. The similar values of critical exponents β, γ, and δ have been also observed for sample B (not shown). Therefore, the Ising-like critical behavior of the specific heat for sample A is inconsistent with the mean-field values for the magnetization exponents. The Ising-like critical behavior of the specific heat has been observed for the itinerant ferromagnet of $SrRuO_3$[13] and it was explained by the mean-field critical behavior, including the (3D) Gaussian fluctuations. Gaussian fluctuations are associated with the variance in M ( $\langle \Delta M^2 \rangle = \langle M^2 \rangle - \langle M \rangle^2$ ).[12] They occur on a short length scale and do not change the mean value of $\langle M \rangle$, therefore, they have no significant effect on magnetization, but give a contribution to $C_p$. The contribution of Gaussian fluctuations to the specific heat is given by $\Delta C = C^{\pm} t^{-\alpha}$, where $\alpha = 2 - d/2$, and d is the dimensionality.[14,15] The amplitude ratio $C^+/C^- = n/2^{d/2}$, where n is the number of spin components. With increasing of Mn concentration the contribution of Gaussian fluctuations to the specific heat of $Ga_{1-x}Mn_xAs$ increases and the critical exponent $\alpha \approx 0.5$ is clearly observed for sample B above and below $T_C$ (see inset of Fig. 4). The value of $C^+/C^- = 0.37$, determined from this experimental plot, is close to the theoretical value of $C^+/C^- = 0.35$ for n = 1, which shows the presence of a strong magnetic anisotropy in the $Ga_{1-x}Mn_xAs$ samples. The Gaussian fluctuation analysis is valid in the same temperature range as mean-field theory and hence has the same Ginzburg criterion for validity: $t > (1 / 32\pi^2) (k_B / \Delta C\ \xi^3)^2$, where $\Delta C$ is the specific heat jump at $T_C$ and $\xi$ is the correlation length.[16,17] Taking $\Delta C$ = 20 J / mole K from Fig.4 and using the smallest t = 0.001 in our experiment yield a lower boundary for the correlation length $\xi > 6.9$ Å.

Very recently, the multifractal spatial variation of the local density of states (LDOS) near the Fermi energy has been observed in GaMnAs by using scanning tunneling microscopy.[18] The LDOS distribution shifts away from Gaussian toward a log-normal distribution with decreasing Mn concentration. However, the log-periodic oscillations in the specific heat of the system with a multifractal distribution is expected,[19] which were not observed in our specific heat measurements. Therefore, our specific heat data support the model of delocalized carriers with a Gaussian distribution, especially in the case of metallic GaMnAs.

In conclusion, we have performed magnetization, resistivity and specific heat measurements on the $Ga_{1-x}Mn_xAs$ to study the critical behavior of the magnetic phase transition in this material. In a close vicinity of the Curie temperature, the temperature dependence of the resistivity and the specific heat reveal the peaks related to the ferromagnetic-paramagnetic phase transition. From detailed analysis of the specific heat data above and below $T_C$, the value of critical exponent $\alpha$ has been determined. The critical behavior of the specific heat of $Ga_{1-x}Mn_xAs$ samples is well described by the mean-field including Gaussian fluctuations model.

## REFERENCES

[1] H. Ohno, Science **281**, 951 (1998).

[2] I. Žutic, J. Fabian, and S. Das Sarma, Rev. Mod. Phys. **76**, 323 (2004).

[3] *Semiconductors and Semimetals,* edited by T. Dietl, D. D. Awschalom, M. Kaminska, and H. Ohno, (Academic Press, 2008) v.82.

[4] T. Dietl, H. Ohno, and F. Matsukura, Phys. Rev. B **63** (2001) 195205.

[5] K. S. Burch, D. D. Awschalom, and D. N. Basov, J. Magn. Magn. Mater. **320** (2008) 3207.

[6] M. E. Fisher, S. –K. Ma, and B. G. Nickel, Phys. Rev. Lett. **29**, 917 (1972).

[7] V. Novák, K. Olejnik, J. Wunderlich, M. Curk, K. Vyborny, A. W. Rushforth, K. W. Edmonds, R. P. Campion, B. L. Gallagher, J. Sinova, and T. Jungwirth, Phys. Rev. Lett. **101**, 077201 (2008).

[8] M. E. Fisher and J. S. Langer, Phys. Rev. Lett. **20**, 665 (1968).

[9] N. O. Birge, P. K. Dixon, and N. Menon, Thermochimica Acta **304/305**, 51 (1997).

[10] Sh. U. Yuldashev, H. Im, V. Sh. Yalishev, C. S. Park, T. W. Kang, S. Lee, Y. Sasaki, X. Liu, and J. K. Furdyna, Appl. Phys. Lett. **82**, 1206 (2003).

[11] E. L. Nagaev, Phys. Rep. **346**, 387 (2001).

[12] S.-K. Ma, *Modern Theory of Critical Phenomena* (Benjamin, New York, 1976), pp. 82-93.

[13] D. Kim, B. L. Zink, F. Hellman, S. McCall, G. Gao, and J. E. Crow, Phys. Rev. B **67**, 100406 (2003).

[14] L. P. Gor'kov, Zh. Éksp. Teor. Fiz. **34**, 735 (1958) [Sov. Phys. JETP **7**, 505 (1958)].

[15] L. G. Aslamazov and A. I. Larkin, Fiz. Tverd. Tela (Leningrad) **10**, 1104 (1968) [Sov. Phys. Solid State **10**, 810 (1968)].

[16] V. L. Ginzburg, Fiz. Tverd. Tela (Leningrad) **2**, 2031 (1960) [Sov. Phys. Solid State **2**, 1824 (1960)].

[17] A. P. Levanyuk, Zh. Éksp. Teor. Fiz. **36**, 810 (1959) [Sov. Phys. JETP **9**, 571 (1959)].

[18] A. Richardella, P. Roushan, S. Mack, B. Zhou, D. A. Huse, D. D. Awshalom, and A. Yazdani : Science **327**, 665 (2010).

[19] I. N. de Oliveira, M. L. Lyra, and E. L. Albuquerque : Physica A **343**, 424 (2004).

**FIGURE CAPTIONS**

Fig. 1. Temperature dependence of the magnetization for (a) 1.6% and (b) 2.6% of $Ga_{1-x}Mn_xAs$ measured at the magnetic field of 10 Oe. Inset: Temperature dependence of the resistivity for (a) 1.6% and (b) 2.6% of $Ga_{1-x}Mn_xAs$ samples at zero magnetic field.

Fig. 2. Temperature dependence of the specific heat for (a) 1.6% and (b) 2.6% of $Ga_{1-x}Mn_xAs$.
The specific heat of GaAs substrate is shown by dash-dotted line.

Fig. 3. Magnetic specific heat of the $Ga_{1-x}Mn_xAs$ ($x = 0.016$) sample. Inset shows the magnetic specific heat versus the reduced temperature $t = |T-T_C| / T_C$ in a double logarithmic scale.

Fig. 4. Magnetic specific heat of the $Ga_{1-x}Mn_xAs$ ($x = 0.026$) sample. Inset shows the magnetic specific heat versus the reduced temperature in a double logarithmic scale.

Fig. 5. M vs H dependence for sample A at T = 40K. Insets: (a) $M^2$ vs H/M and (b) log(M) vs log (H) for sample A at 40K.

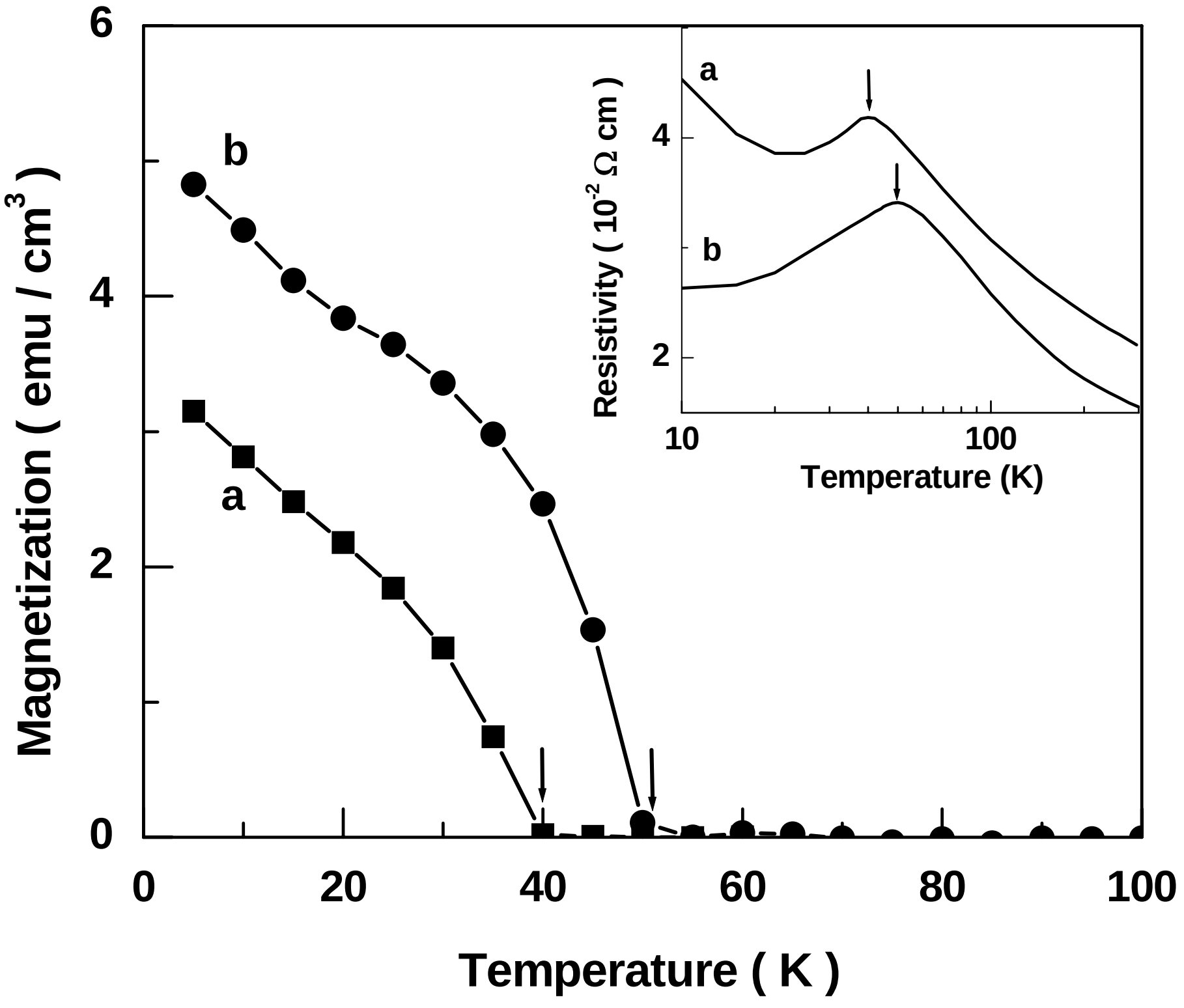

Figure 1

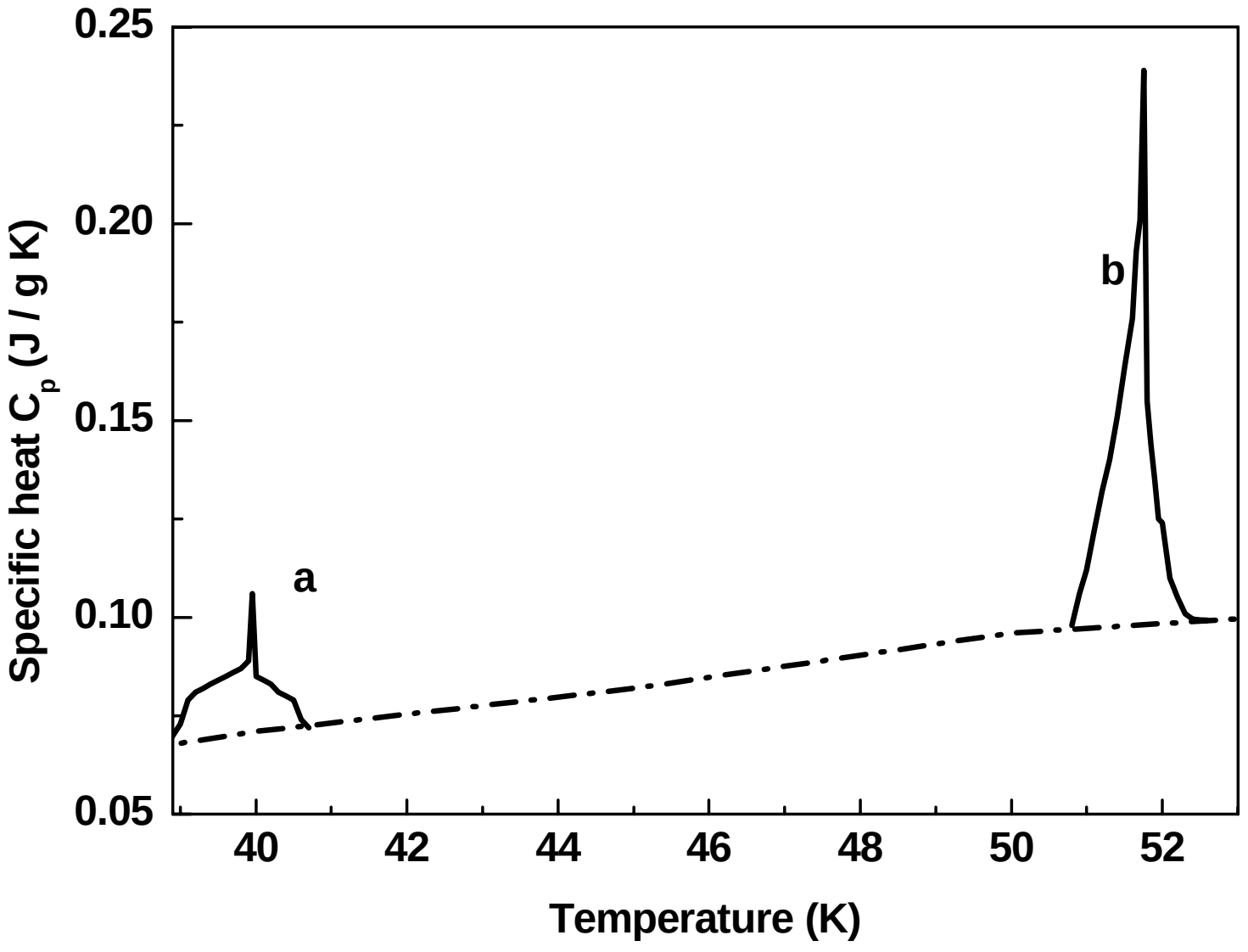

Figure 2

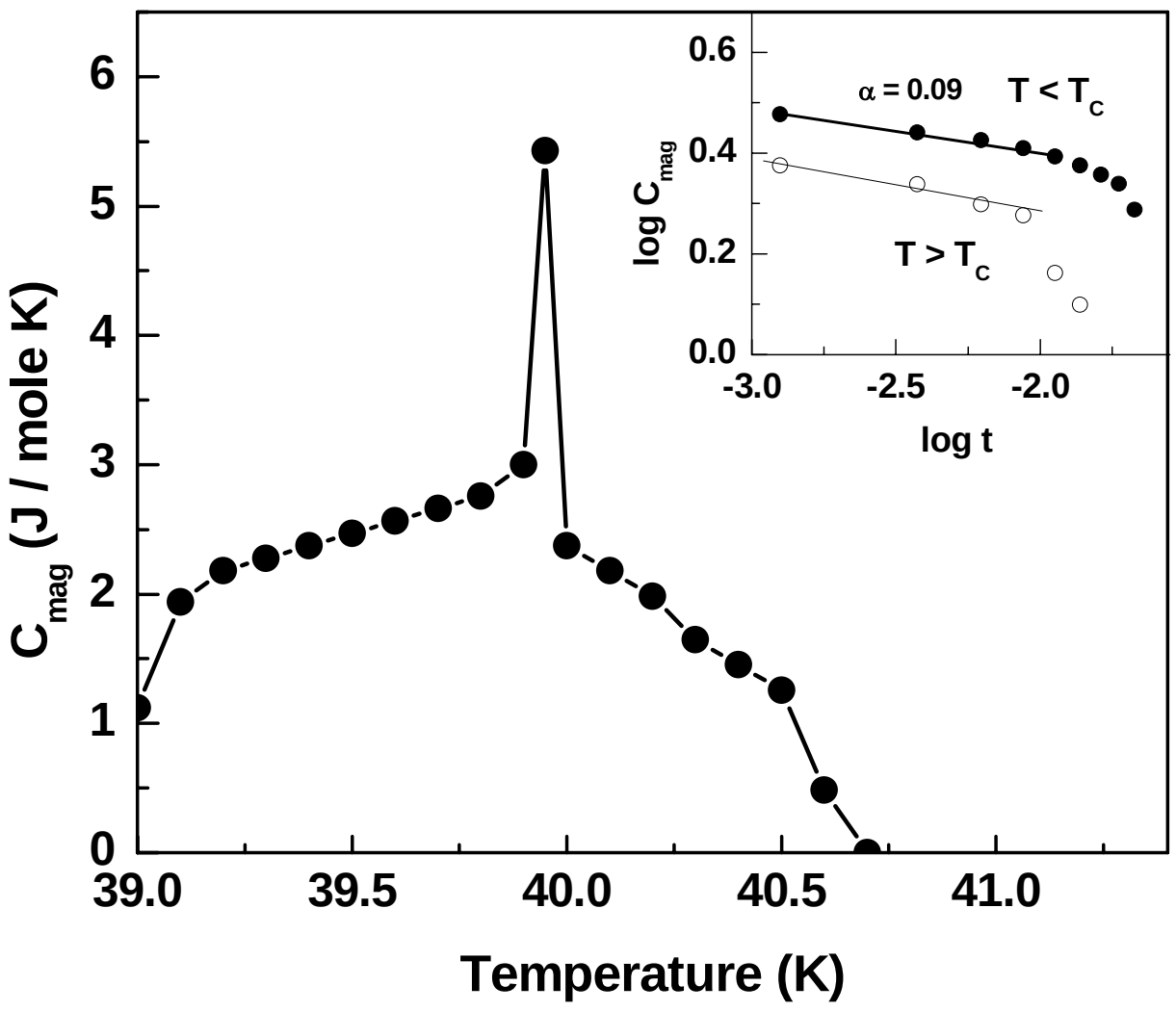

Figure 3

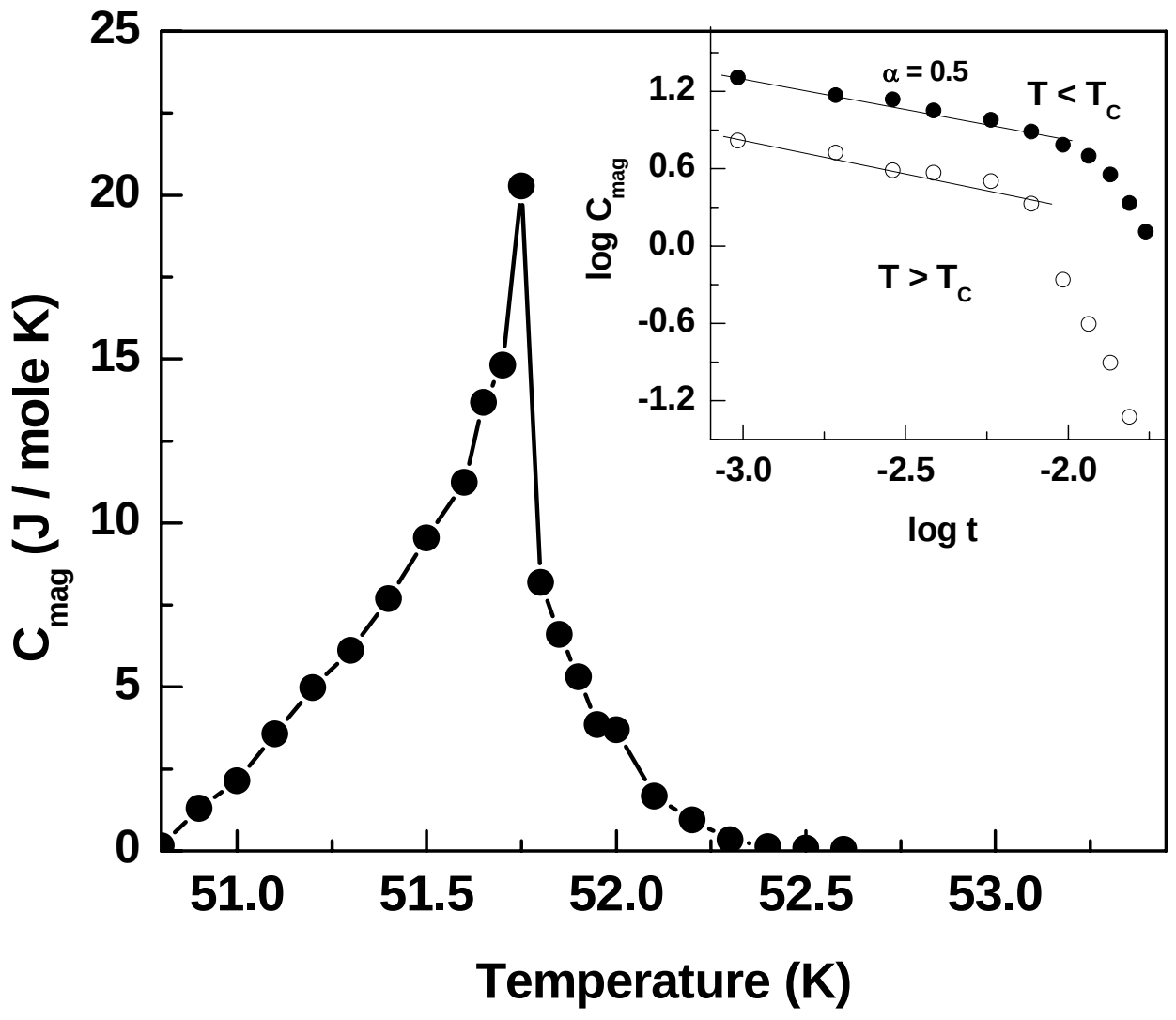

Figure 4

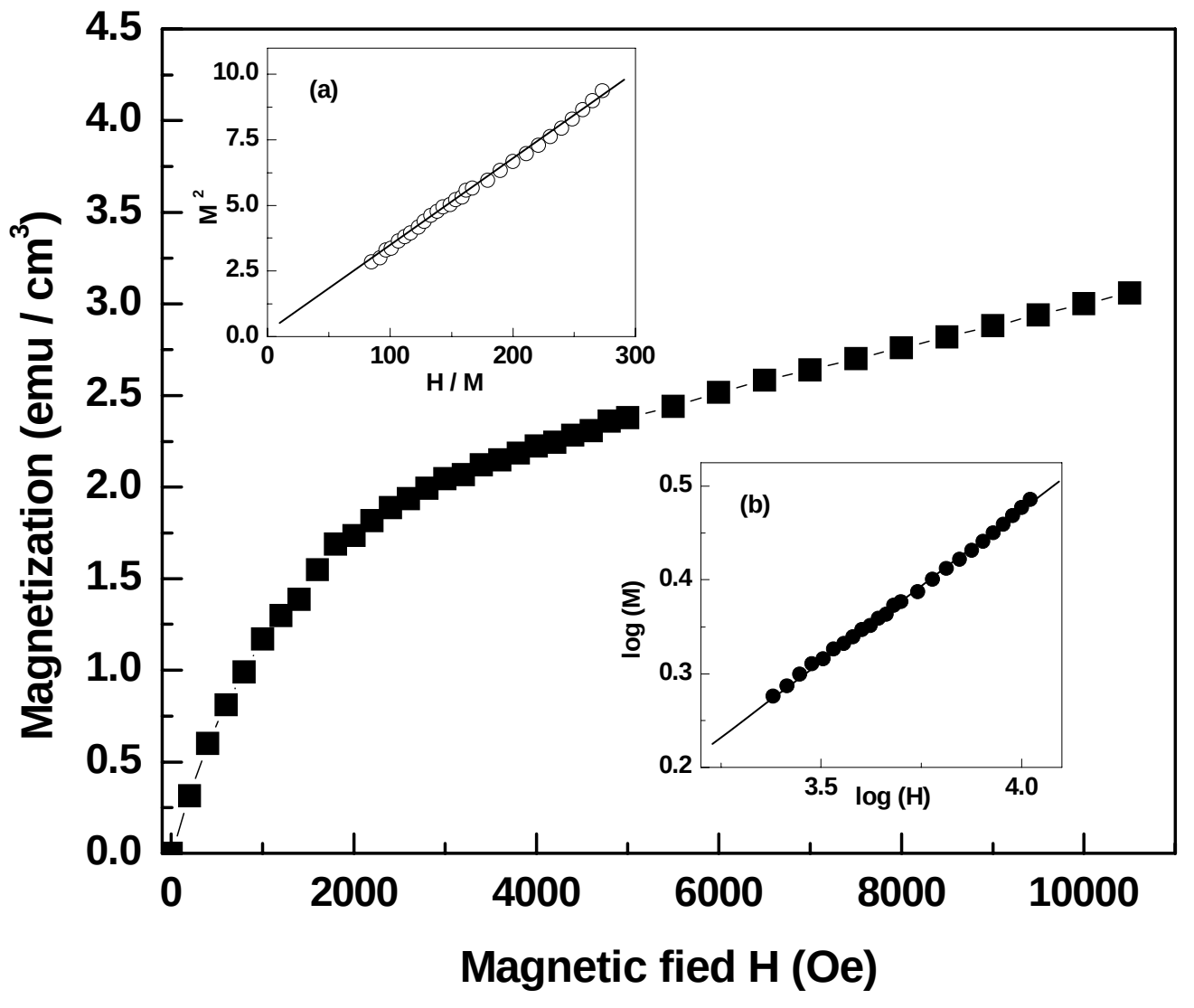

Figure 5